\documentclass[twocolumn]{aastex7}

\begin{document}

\title{Order-by-order Modeling of Exoplanet Radial Velocity Data}

\author[orcid=0000-0001-7574-4440,gname=Zachary, sname=Langford]{Zachary Langford}
\affiliation{Department of Physics and Astronomy, University of Pennsylvania, 209 S 33rd St, Philadelphia, PA 19104, USA}
\email[show]{langfzac@sas.upenn.edu}

\author[orcid=0000-0002-6096-1749,gname=Cullen, sname=Blake]{Cullen Blake}
\affiliation{Department of Physics and Astronomy, University of Pennsylvania, 209 S 33rd St, Philadelphia, PA 19104, USA}
\email[]{chblake@sas.upenn.edu}  

\author[0000-0003-1312-9391]{Samuel Halverson}
\affiliation{Jet Propulsion Laboratory, California Institute of Technology, 4800 Oak Grove Drive, Pasadena, California 91109}
\email{samuel.halverson@jpl.nasa.gov}

\author[0000-0001-6545-639X]{Eric B. Ford}
\affiliation{Department of Astronomy \& Astrophysics, 525 Davey Laboratory, 251 Pollock Road, Penn State, University Park, PA, 16802, USA}
\affiliation{Center for Exoplanets and Habitable Worlds, 525 Davey Laboratory, 251 Pollock Road, Penn State, University Park, PA, 16802, USA}
\affiliation{Institute for Computational and Data Sciences, Penn State, University Park, PA, 16802, USA}
\affiliation{Center for Astrostatistics, 525 Davey Laboratory, 251 Pollock Road, Penn State, University Park, PA, 16802, USA}
\email[]{ebf11@psu.edu}  

\author[0000-0001-9596-7983]{Suvrath Mahadevan}
\affiliation{Department of Astronomy \& Astrophysics, 525 Davey Laboratory, 251 Pollock Road, Penn State, University Park, PA, 16802, USA}
\affiliation{Center for Exoplanets and Habitable Worlds, 525 Davey Laboratory, 251 Pollock Road, Penn State, University Park, PA, 16802, USA}
\affiliation{Astrobiology Research Center, 525 Davey Laboratory, 251 Pollock Road, Penn State, University Park, PA, 16802, USA}   
\email{suvrath@astro.psu.edu}

\author[0000-0002-0078-5288]{Mark R. Giovinazzi}
\affiliation{Department of Physics and Astronomy, Amherst College, 25 East Drive, Amherst, MA 01002, USA}
\email[]{mgiovinazzi@amherst.edu}

\author[0000-0002-5463-9980]{Arvind F.\ Gupta}
\affiliation{U.S. National Science Foundation National Optical-Infrared Astronomy Research Laboratory, 950 N.\ Cherry Ave., Tucson, AZ 85719, USA}
\email{arvind.gupta@noirlab.edu}

\author[0000-0003-0149-9678]{Paul Robertson}
\affiliation{Department of Physics \& Astronomy, The University of California, Irvine, Irvine, CA 92697, USA}
\email{probert1@uci.edu}

\author[orcid=0000-0003-0353-9741,gname=Jaime A.,sname=Alvarado-Montes]{Jaime A. Alvarado-Montes}
\affiliation{Australian Astronomical Optics, Macquarie University, Balaclava Road, Sydney, NSW 2109, Australia.}
\affiliation{Astrophysics and Space Technologies Research Centre, Macquarie University, Balaclava Road, Sydney, NSW 2109, Australia.}
\email{jaime-andres.alvarado-montes@hdr.mq.edu.au}

\author[0000-0003-4384-7220]{Chad F.\ Bender}
\affiliation{Steward Observatory, University of Arizona, 933 N.\ Cherry Ave, Tucson, AZ 85721, USA}
\email{cbender@arizona.edu}

\author[0000-0001-9626-0613]{Daniel M.\ Krolikowski}
\affiliation{Steward Observatory, University of Arizona, 933 N.\ Cherry Ave, Tucson, AZ 85721, USA}
\email{krolikowski@arizona.edu}

\author[0000-0001-8127-5775]{Arpita Roy}
\affiliation{Astrophysics \& Space Institute, Schmidt Sciences, New York, NY 10011, USA}
\email[]{arpita308@gmail.com}  

\author[0000-0002-4046-987X]{Christian Schwab}
\affiliation{School of Mathematical and Physical Sciences, Macquarie University, Sydney, Australia}
\email{mail.chris.schwab@gmail.com}

\author[0000-0002-4788-8858]{Ryan C Terrien}
\affiliation{Carleton College, One North College Street, Northfield, MN 55057, USA}
\email{rterrien@carleton.edu}

\author[0000-0001-6160-5888]{Jason T.\ Wright}
\affiliation{Department of Astronomy \& Astrophysics, 525 Davey Laboratory, 251 Pollock Road, Penn State, University Park, PA, 16802, USA}
\affiliation{Center for Exoplanets and Habitable Worlds, 525 Davey Laboratory, 251 Pollock Road, Penn State, University Park, PA, 16802, USA}
\affiliation{Penn State Extraterrestrial Intelligence Center, 525 Davey Laboratory, 251 Pollock Road, Penn State, University Park, PA, 16802, USA}
\email{astrowright@gmail.com}

\begin{abstract}
Precise radial velocity (RV) measurements are a crucial tool for exoplanet discovery and characterization. Today, the majority of these measurements are derived from Echelle spectra in the optical wavelength region using cross-correlation techniques. Although for certain stars these approaches can produce RVs with sub-1 m~s$^{-1}$ measurement errors, for many others, we are now in a regime where instrumental precision is fundamentally below the intrinsic RV variations of the star that result from astrophysical processes that can be correlated in both time and wavelength. We explore new methods for measuring exoplanet orbital parameters that take advantage of the fact that RV data sets are fundamentally multi-wavelength. By analyzing NEID extremely precise radial velocity (EPRV) data of three known exoplanet systems, we show that fitting a single Keplerian model to multi-wavelength RVs can produce a factor of 1.5 -- 6.8 better $M_p \sin i$ uncertainties compared to fitting RVs that are derived from a weighted average across wavelength.
\end{abstract}

\section{Introduction}\label{sec:intro}
The development of stabilized Doppler spectrometers over the last two decades has revolutionized the study of exoplanets \citep{lovis2010}. These instruments (e.g.,  EXPRES \citep{Blackman2020}, ESPRESSO \citep{Pepe2021}, NEID \citep{schwab2016}, MAROON-X\citep{maroon2018}, Keck Planet Finder \citep{keck2024}, HARPS \citep{harps2003}, and HARPS-N \citep{harpsn} have extended our ability to discover terrestrial planets and, when combined with transit light curves, measure their densities to begin to constrain their compositions \citep[e.g.,][]{leger2009}. Each of these instruments operate across some portion of the optical region (360--1050 nm), employing a similar large-format Echelle grating to generate spectra at a high spectral resolving power of around $R$ =$\lambda / \Delta\lambda$ = 10$^5$ across dozens of individual spectral orders. Measuring precise stellar radial velocities (RVs) requires exquisite calibration of the wavelength scale of the instrument, which is achieved through a combination of wavelength reference sources such as emission line lamps and laser frequency combs, and ensuring that the instrument is intrinsically very stable in terms of temperature and pressure \citep{Robertson2019}. Modern RV spectrometers are, in principle, able to produce stellar RV measurements at the $\sim$ 10 -- 30 cm s$^{-1}$ level, given observations of a sufficient signal-to-noise ratio (SNR) of a star that does not exhibit any astrophysical sources of RV variation \citep[e.g.,][]{Halverson2016}. 

Typically, stellar RV measurements are derived from high-resolution spectra using a cross-correlation technique that has been used with great success for more than 20 years \citep[e.g.,][]{Connes1985,bouchy2001,lovis2010}. The wavelength-calibrated spectra are cross-correlated with a zero-velocity stellar template that encodes information about the locations and relative depths of absorption lines in the stellar spectra of different effective temperatures. The expected photon-noise-limited performance of this technique has been evaluated theoretically and empirically, and depends on the signal-to-noise ratio of the spectra and the number, depths, and widths of the stellar lines. This ``spectral information content," usually denoted $Q$ \citep{bouchy2001}, is higher when there are deeper, narrower, and a greater number of spectral lines --- resulting in better RV precision for a fixed SNR. Ultimately, both the RV estimate and its uncertainty are derived from fitting the peak and width of the cross-correlation function (CCF) between the spectrum and the template. Typically, each of the $N$ spectral orders ($N$ may be up to 100 in some cases) is analyzed separately, producing a set of $N$ contemporaneous RV estimates and uncertainties. Since different spectral orders may have different values of $Q$ or SNR depending on stellar spectral type, these $N$ individual RV estimates must be combined in a weighted sense to generate a single RV estimate representative of the combined spectrum, spanning all $N$ spectral orders. A standard method is to take the weighted mean and variance, using inverse variance weighting, as the RV measurement for a single epoch\footnote{In practice, the computation is slightly different in some pipelines. The NEID DRP, for example, fits a weighted sum of the $N$ CCFs with a single Gaussian. The weights take into account the throughput of the instrument and the spectral information in a given order, and are fixed across time for a single target.}. For the purposes of this paper, we call this the Variance-Weighted Mean (VWM) RV method. The VWM method is widely used and has been shown to produce sub-1 m s$^{-1}$ RVs in certain situations \citep{proxima2020}. Alternative methods involving full forward modeling of the spectra or template matching are also used in the literature, but are often found to produce only modest improvements in RV precision at optical wavelengths for sun-like stars, despite a significantly higher computational burden. For cooler stars with more complex spectra, the forward-modeling approach can be superior to the CCF approach \citep{Anglada2012, SERVAL2017, Silva2022}.

A key limitation of the VWM method is that it is only an optimal way to estimate the overall stellar RV in the limit where the uncertainties in the N spectral order RV measurements are dominated by photon noise and spectral line shapes do not change. For example, a certain spectral order my have a high SNR (many photons per pixel detected) and a high Q (numerous deep and narrow spectral lines), but also contain low-level chromospheric emission features that vary in time, representing a significant source of correlated RV noise. Or, a certain spectral order may be contaminated by low-level telluric lines that beat against features in the stellar spectrum as the barycentric velocity changes, resulting in spurious RV signals \citep{sharon2022}. The traditional cross-correlation-based VWM method might assign this problematic spectral order a large weight in the final weighted combination of N individual RV estimates, producing a biased overall stellar RV measurement. 

While the Keplerian signal describing the reflex motion of the host star is intrinsically achromatic, various instrumental and astrophysical sources of RV noise are not. Previous work has analyzed RVs in a wavelength-dependent or line-by-line way \citep[e.g.,][]{Xavier2018, artigau2022, Moulla2022, Burrows2024, moulla2024}. In these approaches, the spectral subregions do not have to be full spectral orders and could instead be individual spectral lines, resulting in a very large N. In some cases, these line-by-line approaches produce superior RV measurements and also allow for the association of stellar noise with specific astrophysical processes, such as depth of line formation within the stellar atmosphere \citep{moulla2024}. 

In this paper, we explore different methods for measuring orbital parameters of exoplanet systems using multi-order spectroscopic data (Section \ref{sec:methods}). We show that these methods can recover orbital parameters with greater precision than the VWM method for a set of publicly available NEID data for known exoplanet systems (Section \ref{sec:app}). These modeling paradigms are straightforward to implement and our implementation relies entirely on existing code bases, namely \texttt{Octofitter.jl}\footnote{\url{https://github.com/sefffal/Octofitter.jl}}\citep{thompson2023a}. We provide the Julia code used in our analyses as examples in a public open-source GitHub repository\footnote{\url{https://github.com/langfzac/obo-paper}}.

\section{Modeling Paradigms}\label{sec:methods}
In this section, we outline the three modeling paradigms that we use in our analysis. Section \ref{ssec:vwm} describes the Variance-weighted mean (VWM) method -- a standard method used in the field. Sections \ref{ssec:obo} and \ref{ssec:jk} describe the new methods we are exploring in this work. VWM in essence is simply a way of reducing the data before applying some probabilistic model and, generally, sampling the posteriors with Markov-chain Monte Carlo (MCMC). With two new methods, we aim to better incorporate the information in each of the individual spectral orders' RVs into the final measurements of the orbital parameters.

We employ the typical Keplerian orbit RV model for each of our methods, which has five parameters: orbital period $P$, eccentricity $e$, time of periastron passage $t_p$, argument of periastron $\omega$, and RV semi-amplitude $K$. The exoplanet minimum mass, $M_p\sin{i}$, is derived using Kepler's Laws from the orbital parameters and an assumed stellar mass, $M_*$.

For all log-likelihood computations, we use a Gaussian log probability density function (as implemented in \texttt{Octofitter.jl}):\\
\begin{eqnarray}\label{eqn:likelihood}
    \mathcal{L}{(\{\text{rv}_i\} \mid \vec{\theta},\gamma,\sigma_{\text{jitter}})} = -\frac{M}{2}\log{(2\pi(\sigma_i^2+\sigma_{\text{jitter}}^2))} \nonumber \\ - \frac{1}{2} \sum_{i=1}^{M} \frac{\left(\text{rv}_i - \mu_{\text{rv}}(\vec\theta, t_i) - \gamma \right)^2}{\sigma_i^2 + \sigma_{\text{jitter}}^2}
\end{eqnarray}
where $\text{rv}_i$ is the RV measurement at epoch $t_i$, $\sigma_i$ is the uncertainty associated with a given $\text{rv}_i$, $M$ is the total number of epochs for the set of measurements $\{\text{rv}_i\}$, $\vec{\theta}$ are the set of Keplerian parameters, $\sigma_{\text{jitter}}$ is an additional white-noise term (``jitter"), $\gamma$ is a linear RV offset, and $\mu_{\text{rv}}$ represents the Keplerian RV model evaluated at the epoch $t_i$.

\subsection{Variance Weighted Mean}\label{ssec:vwm}
The VWM method is commonly used to fit Keplerian orbits to RV data. This involves measuring the Doppler shift for each spectral order and combining them through a weighted mean. The weights are often taken to be the inverse variance for each order's RV measurement, based on the photon-noise-limited RV uncertainty. The typical weighted mean and variance, assuming inverse-variance weights, are calculated:
\begin{eqnarray}\label{eqn:vwm}
    &\bar{\text{rv}} &= \frac{\sum_{i=1}^{N} \text{rv}_i / \sigma_i^2}{\sum_{i=1}^{N} 1/\sigma_i^2}, \\
    &\bar{\sigma}^2 &= \frac{1}{\sum_{i=1}^{N}\sigma_i^2}.
\end{eqnarray}
The underlying assumption in this method is that each order's RV measurement is uncorrelated (i.e., the measured Doppler shift is achromatic).

Once a VWM RV is obtained for each epoch, we fit the resulting time-series with a Keplerian orbit and derive posterior distributions for the parameters via MCMC. We use a Gaussian log-likelihood function, as shown in Equation \ref{eqn:likelihood}, which assumes the noise is simply the given uncertainties and some (free-parameter) additional white-noise, uniform across all epochs. Since VWM is common practice, we use this as a baseline against which to compare our new methods.

\subsection{Order-by-Order}\label{ssec:obo}
The first new method we explore is simply to compute a posterior distribution over each Keplerian orbital parameter for each of the $N$ spectral orders' time-series, individually. From these $N\times$5 posteriors, we estimate $N$ means and variances for each Keplerian orbital parameter. We then take the weighted mean and variance (using Equation \ref{eqn:vwm}) for each parameter to arrive at a final measurement. In the limit of white noise (and high SNR), this method should produce the same result as the VWM method. In this limit, we would expect each orders' measurement to be scattered about the mean in an unbiased way. Any deviations would indicate there may be correlated noise or other effects.

We use the same log-likelihood function as in the VWM case (Equation \ref{eqn:likelihood}) for each of the $N$ spectral orders -- a total of seven free-parameters including the offset and jitter terms, per order. This paradigm attempts to account for any additional noise contained in any particular spectral order (i.e., chromatic noise) that is uniform across the time-series. This can also serve as a diagnostic tool. Since each order is fit with a distinct Keplerian model, if an achromatic RV signal is present, the high SNR or uncontaminated orders should produce consistent results. Here, we have shifted the assumption of uncorrelated measurements from the individual order RVs to the final measurements of the parameters. We refer to this method as ``Order-by-Order" (OBO).

\subsection{Joint Keplerian}\label{ssec:jk}
The second new method we test is to fit a single Keplerian model jointly to all the data across the $N$ spectral orders at once. We still allow each order to have its own jitter and offset, but sample for a single set of Keplerian orbit parameters: $2\times N + 5$ total free parameters. The log-likelihood function becomes: 
\begin{equation}
    \mathcal{L} = \sum_{j=1}^{N}\mathcal{L}{(\{\text{rv}_i\}_{j} \mid \vec{\theta}, \gamma_j, \sigma_{\text{jitter},j})}.
\end{equation}
Similar to the OBO paradigm, we expect that if an achromatic signal is present, it will be present across all orders. This method makes this assumption explicit in the probabilistic model. That is, the posterior probability of the model parameters is now jointly conditioned on all of the data at once. This method now attempts to account for any wavelength correlations that are uniform across time -- the jitter and offset terms are now sampled jointly. Of the three modeling schemes, this is the best motivated probabilistic model for these data. We refer to this method as ``Joint Keplerian" (JK).

\section{Applications} \label{sec:app}
In this section, we apply each modeling scheme to three publicly available data sets (Section \ref{ssec:data}) from the NEID instrument -- a high-precision Echelle spectrograph installed at the 3.5m WIYN telescope\footnote{The WIYN Observatory is a joint facility of the NSF’s National Optical-Infrared Astronomy Research Laboratory, Indiana University, the University of Wisconsin-Madison, Pennsylvania State University, Purdue University and Princeton University.}. We carry out Bayesian analysis of these data using MCMC (Section \ref{ssec:mcmc}) for each of our modeling paradigms. We discuss the qualitative properties of the resulting $M_p \sin(i)$ marginal posterior distributions for each method (Section \ref{ssec:msini}), and of the individual OBO posteriors (Section \ref{ssec:obo_msini}). Finally, we describe the overall results and the computational performance of each method (Section \ref{ssec:results}), and compare our results for HD 217107 to a recent analysis of the same data (Section \ref{ssec:mark}).

\subsection{Data}\label{ssec:data}
To ensure our comparison is as constrained as possible, we only use data for systems that meet the following criteria: 1) previously published orbital parameter measurements\footnote{HD 217107 has a confirmed planet on a $\sim$5000 day orbital period. We model the companion with a linear trend, which adds a single ``slope" parameter. In our testing, this did not meaningfully effect the precision of the parameter measurements, but does slightly shift the mean posteriors toward lower mass, as expected.}, 2) have at least 20 epochs covering most of the orbital phase, and 3) are contained in a single observing run, so that we may ignore long-term systematics such as instrumental offsets \citep{Gupta2025b}. This leaves us with three systems that span a range of RV semi-amplitudes with $K$ = 141, 16, and 2 m~s$^{-1}$, and a range of orbital periods $P$ = 7, 62, and 31 days, respectively. Table \ref{tab:data} shows the previously reported parameter measurements (``Literature"), the measurements we obtain using the JK method (``This work"), the total number of observations, the total number of ``valid" spectral orders (see below for discussion), the start date, end date, and total time baseline of the observations, and the median error of the VWM RVs. We are not making a direct comparison to the literature measurements, but we include the values to check that our measurements are reasonable.

Each data set we use is reduced from publicly available NEID data\footnote{\url{https://neid.ipac.caltech.edu/search.php}}, which were collected as part of the NEID Earth Twin Survey \citep{gupta2021}. We use the Level 2 data products from the NEID Data Reduction Pipeline (DRP) version 1.4.0\footnote{\url{https://neid.ipac.caltech.edu/docs/NEID-DRP/}}, which provide an RV measurement for each Echelle order at a given observation time (the \texttt{CCRV} keyword in the FITS header), along with the full CCF for each order. We estimate the statistical uncertainty on the RV estimate for each order from the width of the CCF following \citet{bouchy2001}. We remove any data points that are flagged as having corrupted RVs or CCFs (e.g., RV=0.0 m s$^{-1}$ and $\sigma_{RV}=10^9$ m s$^{-1}$) for a particular spectral order. We then remove the corresponding data point across all epochs. This ensures that there are no additional effects from some orders having more data points than others. We shift each of the individual order time-series to zero RV by subtracting the median across all epochs\footnote{This makes setting a prior on the linear offset simpler, as it can then be take as a distribution that is centered at zero for all systems.}. The final data set for each system is comprised of an RV measurement and its uncertainty, for each Echelle order that passes the quality cuts, at each epoch.

For the VWM analyses, we compute the inverse-variance weighted mean and variance (see $\S$ \ref{ssec:vwm}) of the order-by-order RV measurements at each epoch. In this case, the linear shift to the median RV is carried out on the weighted mean RVs, rather than each order's time-series.  

\begin{table*}
\begin{center}
\begin{tabular}{cc|ll|ll|ll}%
\tableline
& & \multicolumn{2}{c}{HD 217107\tablenotemark{a}} & \multicolumn{2}{c}{HD 3651\tablenotemark{a}} & \multicolumn{2}{c}{HD 86728\tablenotemark{b}} \\ 
Parameter & Units & Literature & This work & Literature & This work & Literature & This work\\ \tableline
$M_*$ & $M_\odot$ & 1.060 & -- & 0.888 & -- & 0.967 & -- \\
$M_p\sin{i}$ & $M_{\text{jup}}$ & 1.385 $\pm$ 0.039 & 1.389 $\pm$ 0.0017 & 0.220$^{+0.0078}_{-0.0076}$ & 0.228$^{+0.0019}_{-0.0017}$ & 0.030 $\pm$ 0.0018 & 0.0275$^{+0.0024}_{-0.0023}$ \\
$K$ & m~s$^{-1}$ & 141.91 $\pm$ 0.26 & 142.34$^{+0.17}_{-0.18}$ & 15.50$^{+0.39}_{-0.41}$& 15.79$^{+0.14}_{-0.13}$ & 1.97 $\pm$ 0.12 & 1.88$^{+0.16}_{-0.17}$ \\
$P$ & days & 7.126$^{+0.0000067}_{-0.0000065}$ & 7.127$^{+0.00019}_{-0.00016}$ & 62.242 $\pm$ 0.0032 & 61.94$^{+0.08}_{-0.07}$ & 31.151$^{+0.0059}_{-0.0066}$ & 30.0$^{+0.7}_{-0.8}$ \\
$e$ & -- & 0.128$^{+0.0017}_{-0.0016}$ & 0.127$^{+0.0011}_{-0.0012}$ & 0.614 $\pm$ 0.015 & 0.598 $\pm$ 0.006 & 0.067$^{+0.065}_{-0.046}$ & 0.12$^{+0.15}_{-0.08}$ \\
$\omega$ & deg & 22.6 $\pm$ 0.8 & 204.3 $\pm$ 0.4 & 63.21 $\pm$ 2.10 & 61.0$^{+1.0}_{-1.1}$ & 92.0$^{+43.0}_{-113.0}$ & 9$^{+47}_{-81}$ \\
Epochs &  & & 22 & & 53 & & 20 \\
Orders &  & & 77 & & 76 & & 77 \\
Start Date & JD & & 2459474.79 & & 2459894.79 & & 2459914.89 \\
End Date & JD & & 2459738.95 & & 2460194.88 & & 2460014.70 \\
Baseline & days & & 264.2 & & 300.1 & & 99.8 \\
Median Error & m~s$^{-1}$ & & 0.289 & & 0.332 & & 0.360 \\
\tableline
\end{tabular}
\end{center}
\tablenotetext{a}{\cite{rosenthal2021}}
\tablenotetext{b}{\cite{Gupta2025a}}
\caption{Keplerian orbital parameters for each system as given on the Exoplanet Archive (``Literature") and those computed by our Joint Keplerian (JK) analysis (``This work"; Figures \ref{fig:pairplot_HD217107}, \ref{fig:pairplot_HD3651}, and \ref{fig:pairplot_HD86728}), total number of included observations, total number of spectral orders per epoch, the start date, end date, and total time baseline of the observations, and the median error across the time-series for the VWM data. We leave out $t_p$ since the conventions in the literature are inconsistent. We note that the literature parameters are derived from analyses using many more data points than we use in this work.}
\label{tab:data}
\end{table*}

\subsection{MCMC Sampling}\label{ssec:mcmc}
The specific aim of our tests is to see how well we can estimate the minimum planet mass, $M_p\sin{i}$. In an attempt to minimize any additional differences in the modeling, we choose a standard set of priors on the Keplerian parameters for each system. Since all of the systems have reported values in the literature\footnote{For HD 86728, the discovery paper \citep{Gupta2025a} includes the analysis of the data used in this paper (along with additional data). Because of this, our analysis of the system, which includes prior distributions informed by that paper, is no longer a proper Bayesian analysis. However, the goal of this work is simply to demonstrate a promising new method, not to provide new, robust measurements of the parameters.}, we treat each analysis as a quasi-Bayesian updating problem. In all cases, we fix the stellar mass when we infer $M_p\sin{i}$; we place a wide Gaussian prior on $K$ ($\mu$ = 0, $\sigma$ = 500; (m~s$^{-1}$)), truncated at $K$ $<$ 0 m~s$^{-1}$; the period has a Gaussian prior centered on the literature measurement, with a width of 1.5 days; eccentricity has a Gaussian prior centered on the literature value, with a width of five times the reported eccentricity uncertainty and truncated at 0; the argument of periastron is given a uniform prior on [0, 2$\pi$]; the time of periastron passage is given a uniform prior which spans an orbital period\footnote{In the actual sampling, this is derived from a uniform distribution on [0,1] times the period in days.}; the jitter terms are given a wide Gaussian prior ($\mu$=0, $\sigma$=25.0; (m~s$^{-1}$)), truncated at values below 0; and the $\gamma$ offset terms have a Gaussian prior with $\mu$ = 0 and $\sigma$ = 1000 (m~s$^{-1}$). For the OBO method, we use the same priors for each order, and for the JK method, we use the same priors on $\gamma$ and $\sigma_{\text{jitter}}$ for each order. 

The MCMC analysis is carried out using an implementation of the No-U-Turn Sampler \citep[NUTS;][]{hoffman2014}. We use the default \texttt{Octofitter.jl} sampling setup, which starts by finding an estimate of the maximum posterior parameters and covariance matrix to initialize NUTS. We tune the NUTS sampling parameters (leapfrog integration stepsize, number of steps, and the so-called 'mass-matrix') using 1000 adaptation steps, after which we sample for 1000 iterations. For the OBO case, we found that some of the models produce chains with large correlation lengths. To ensure we obtain an accurate representation of the posterior, if the autocorrelation length of any parameter chain at the first step is greater than 0.5 we continue sampling until we have at least 500 effective samples. We use the implementation in \texttt{MCMCChains.jl}\footnote{\url{https://github.com/TuringLang/MCMCChains.jl}} to compute the number of effective samples and the auto-correlation function for each parameter. The analyses in this paper were carried out on a computing node with two AMD EPYC 7543 32-Core processors and 250 GB of memory and \texttt{Julia} version 1.10.9 for x86\_64 Linux. 

\subsection{$M_p\sin{i}$ Marginal Posteriors}\label{ssec:msini}
\begin{figure*}[t]
    \gridline{
        \fig{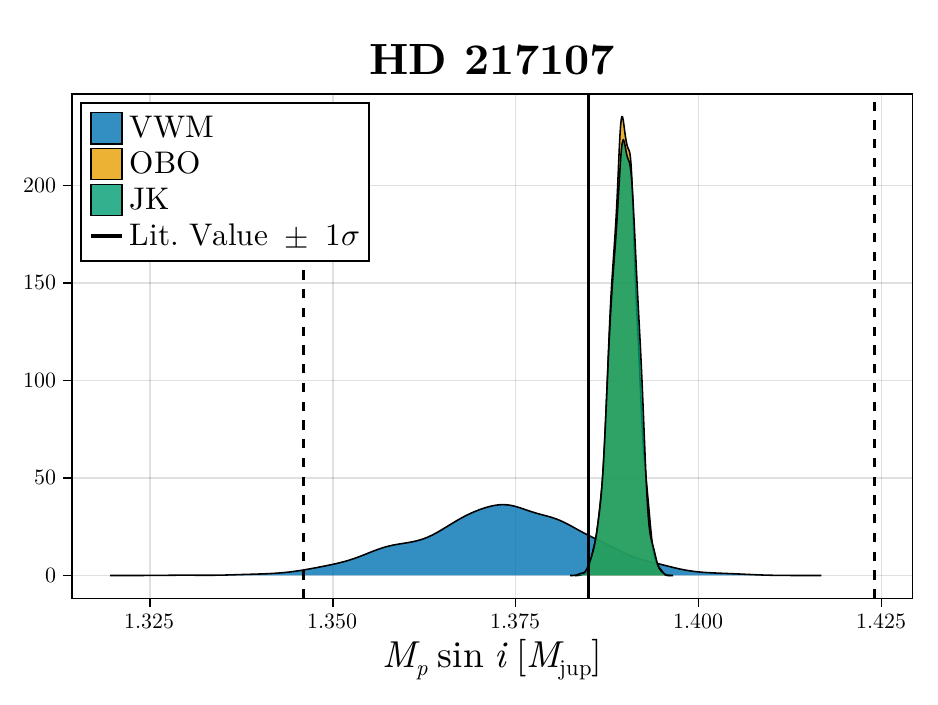}{0.5\textwidth}{(a)}
        \fig{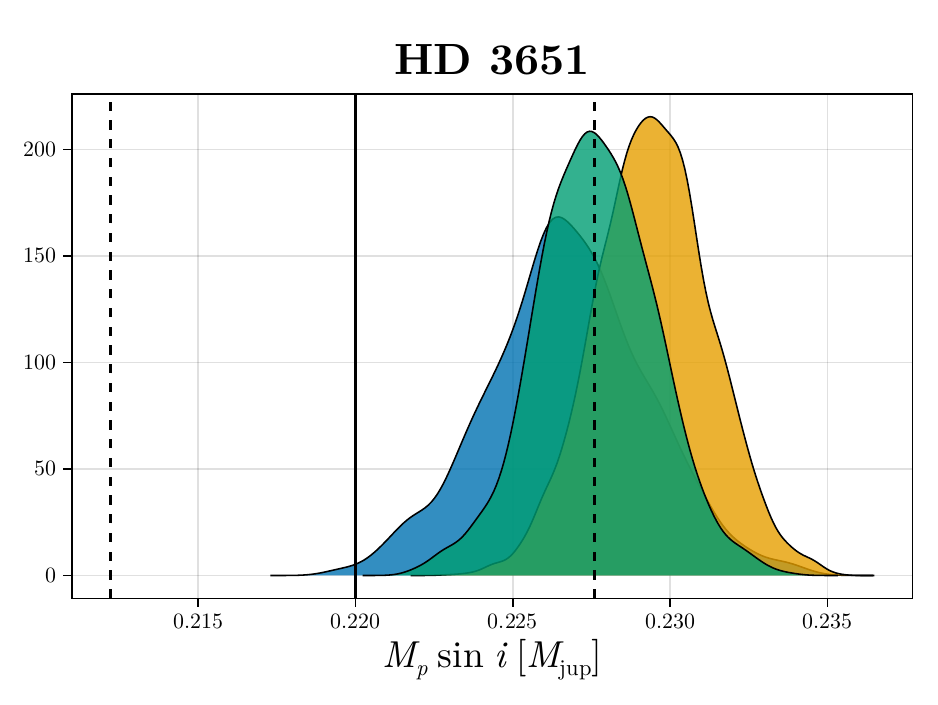}{0.5\textwidth}{(b)}}
    \gridline{
        \fig{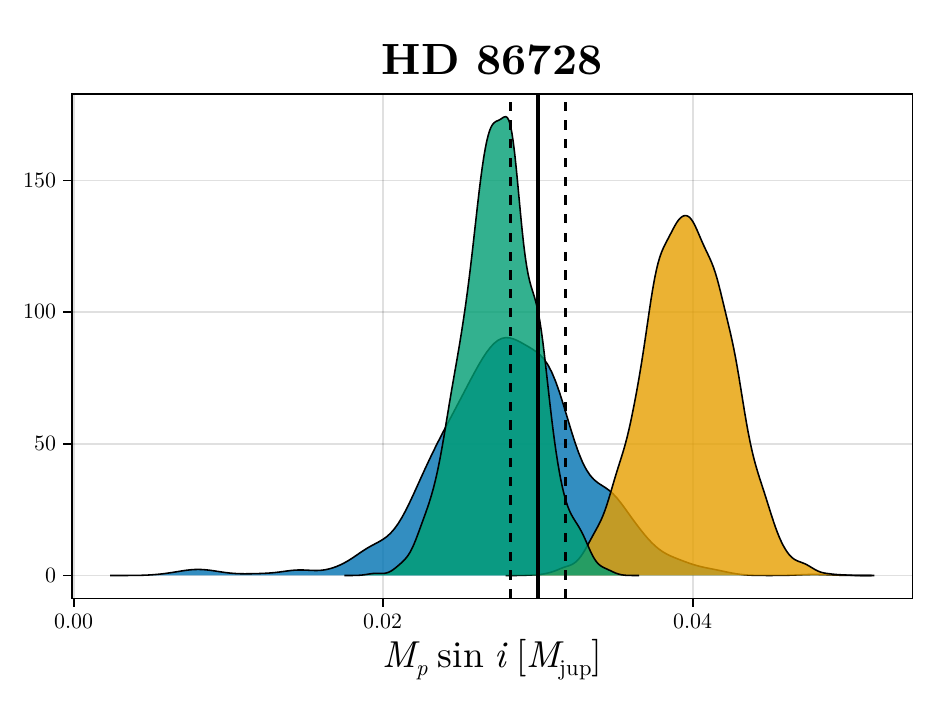}{0.5\textwidth}{(c)}
        }
    \caption{Marginal posterior probability distributions of $M_p\sin i$ given the data for each system, for each of the three modeling paradigms. The blue represents the Variance-weighted-mean method, orange is the Order-by-order, green is the Joint Keplerian method, and the black lines show the mean (solid) and 1-sigma uncertainties (dashed) reported in the literature. Note in panel a), the OBO and JK methods are nearly indistinguishable, visually.}
    \label{fig:mass_posteriors}
\end{figure*}

The marginal posteriors on $M_p\sin{i}$ for each method are shown in Figure \ref{fig:mass_posteriors}, based on the inferred orbital parameters and a fixed stellar mass. The blue densities are VWM, the orange are OBO\footnote{For these figures, we compute the mean and variance of each order's posterior samples. Each combination is treated as an independent measurement of the parameter. We then compute the inverse-variance weighted mean and variance for the parameter. The figure shows 5000 samples from a Gaussian described by those values.}, the green are JK, and the black lines represent the literature reported measurement (solid) and uncertainties (dashed). In all cases, the VWM model provides a less precise measurement of $M_p\sin{i}$  than OBO or JK. The RV semi-amplitudes of our systems span a range of nearly a factor of 100. We find the system with the largest RV semi-amplitude (HD 217107) exhibits the most significant improvement in parameter uncertainties with the OBO or JK methods compared to the VWM method (see Fig~\ref{fig:mass_posteriors}). The systems HD 3651 and HD 86728 have substantially smaller RV semi-amplitudes than HD 217107 and exhibit less significant improvements in parameter uncertainties in our analyses. This suggests that there may be a relationship between the strength of the planetary signal and the performance of the OBO or JK methods. At the same time, we see that the bias between the VWM and OBO and JK parameter estimates is largest for HD 217107. The relationships between the benefits of the JK and OBO methods and system characteristics will be explored in future work.

\subsection{Order-by-order $M_p\sin{i}$}\label{ssec:obo_msini}
\begin{figure*}
    \gridline{
        \fig{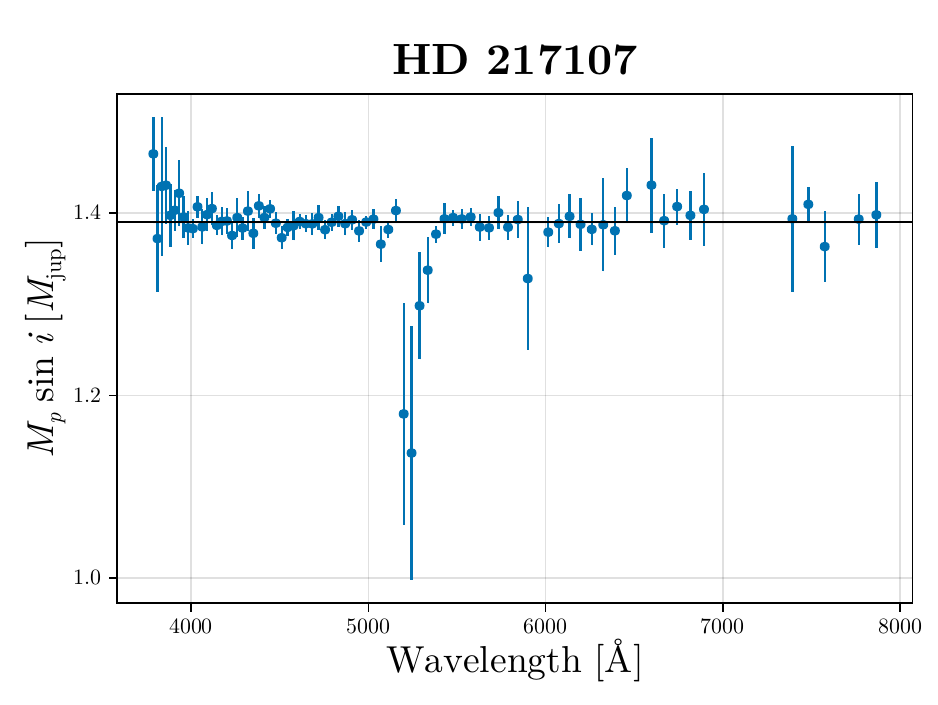}{0.5\textwidth}{(a)}
        \fig{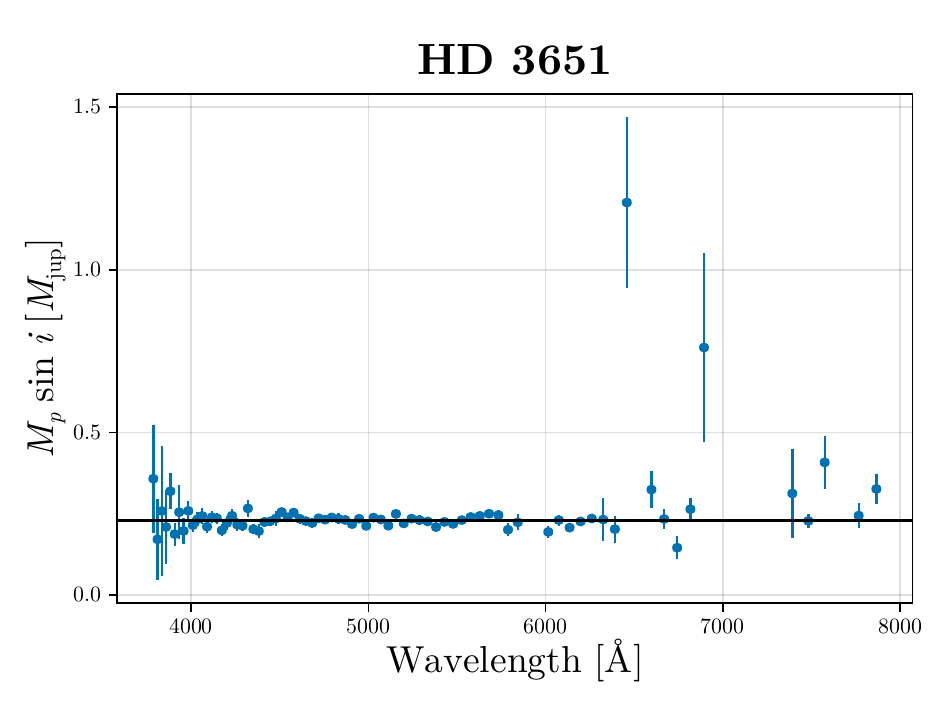}{0.5\textwidth}{(b)}}
    \gridline{
        \fig{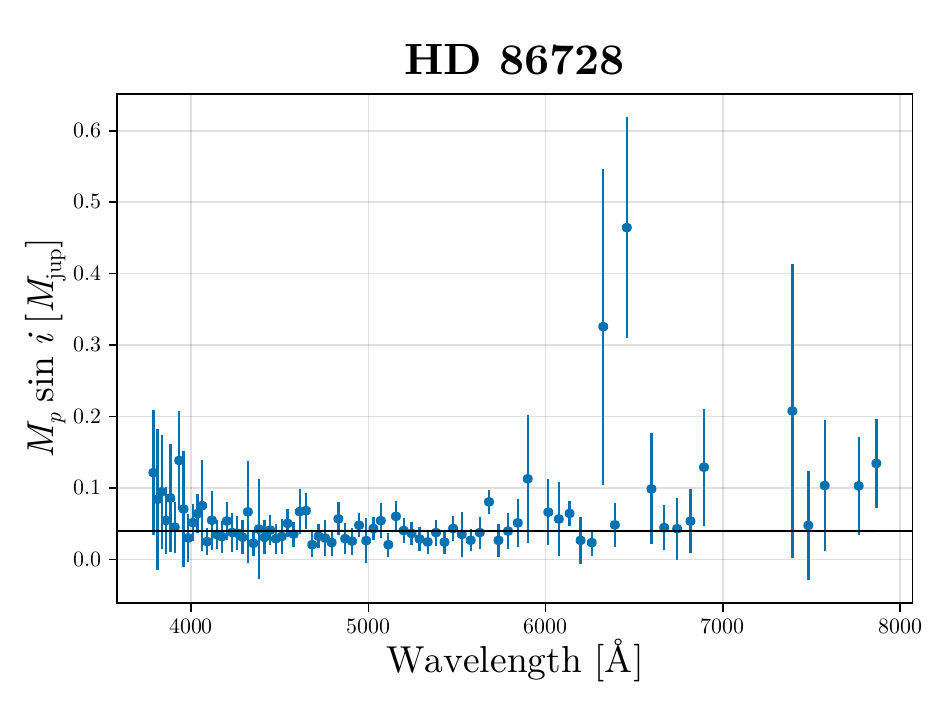}{0.5\textwidth}{(c)}
        }
    \caption{Measurements of $M_p\sin{i}$ derived from each spectral order's individual MCMC analysis as a function of central wavelength. Blue points and errors are the mean and standard deviation of the posterior distributions for each order. The solid black line represents the variance weighted mean of the blue points.}
    \label{fig:obo_masses}
\end{figure*}

In this section we examine the $M_p\sin{i}$ measurements as a function of spectral order. Figure \ref{fig:obo_masses} shows the mean (blue points) and standard deviation (blue segments) of the $M_p\sin{i}$ marginal posterior distribution of each order, for the three exoplanet systems that were analyzed. The black line represents the inverse-variance weighted mean of the blue points. We can see that there are some wavelength regions that show large variations from the mean (black line), which are common to each data set. For example, at the blue end ($\lesssim$4000 Angstroms), there is a clear upward trend as the wavelength decreases. In addition, we see that there are a number of orders in each data set that significantly vary in their mean values, some of which are biased by many multiples of their computed uncertainties.

\begin{figure*}
    \gridline{
        \fig{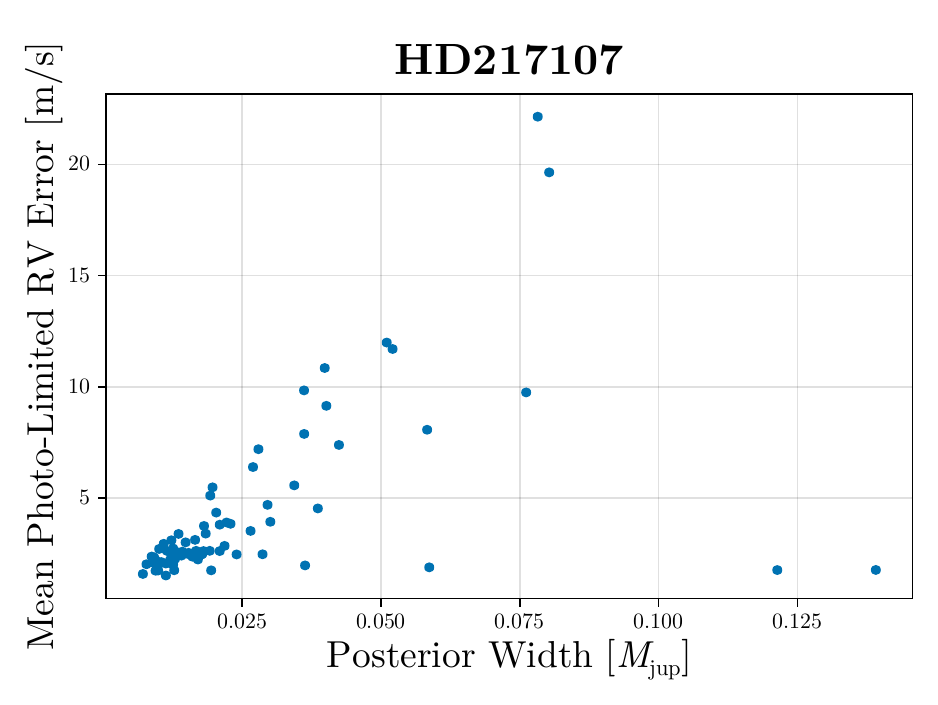}{0.5\textwidth}{(a)}
        \fig{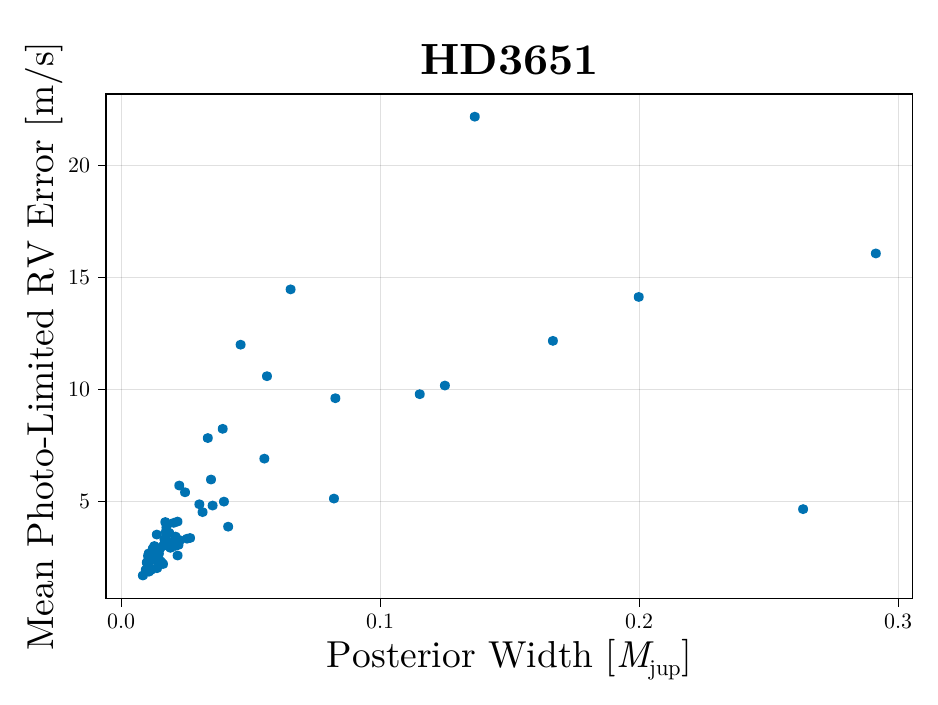}{0.5\textwidth}{(b)}}
    \gridline{
        \fig{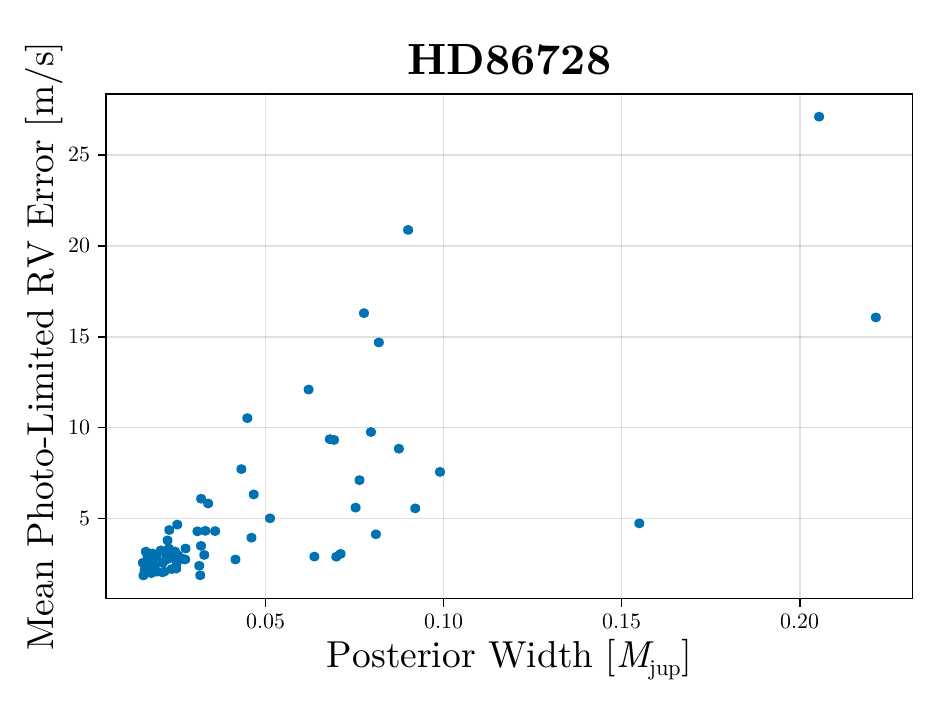}{0.5\textwidth}{(c)}
        }
    \caption{Mean of the photon-limited uncertainties versus the standard deviation of the $M_p\sin{i}$ posteriors for each order. Each blue point represents one spectral order. This figure shows a loose correlation between the average uncertainty in the RV time-series (for a single order) and the resulting constraint on $M_p\sin{i}$. Orders with small formal RV uncertainties can produce Keplerian orbital parameters that are poorly constrained. }
    \label{fig:rms}
\end{figure*}

Figure \ref{fig:rms} shows the mean of the photon-limited uncertainties versus the standard deviation of the $M_p\sin{i}$ posteriors for each order. Each blue point represents one spectral order. This illustrates that there is only a loose correlation between the average uncertainty in the RV time-series (for a single order) and the resulting constraint on $M_p\sin{i}$. That is, we can see that some orders that have small RV uncertainties can still have posteriors with relatively large widths. A large mean RV error and small posterior width might signify that the RV errors are overestimated, possibly because the CCF is being incorrectly fit. A small mean RV error and a large posterior width might suggest that significant systematic error due to stellar activity results in errors that are effectively underestimated. Regardless, it is clear that there is some degree of chromatic noise present in these data.

\subsection{Joint Posteriors}
\begin{figure*}
    \centering
    \textbf{HD 217107}\par\medskip
    \includegraphics[width=\textwidth]{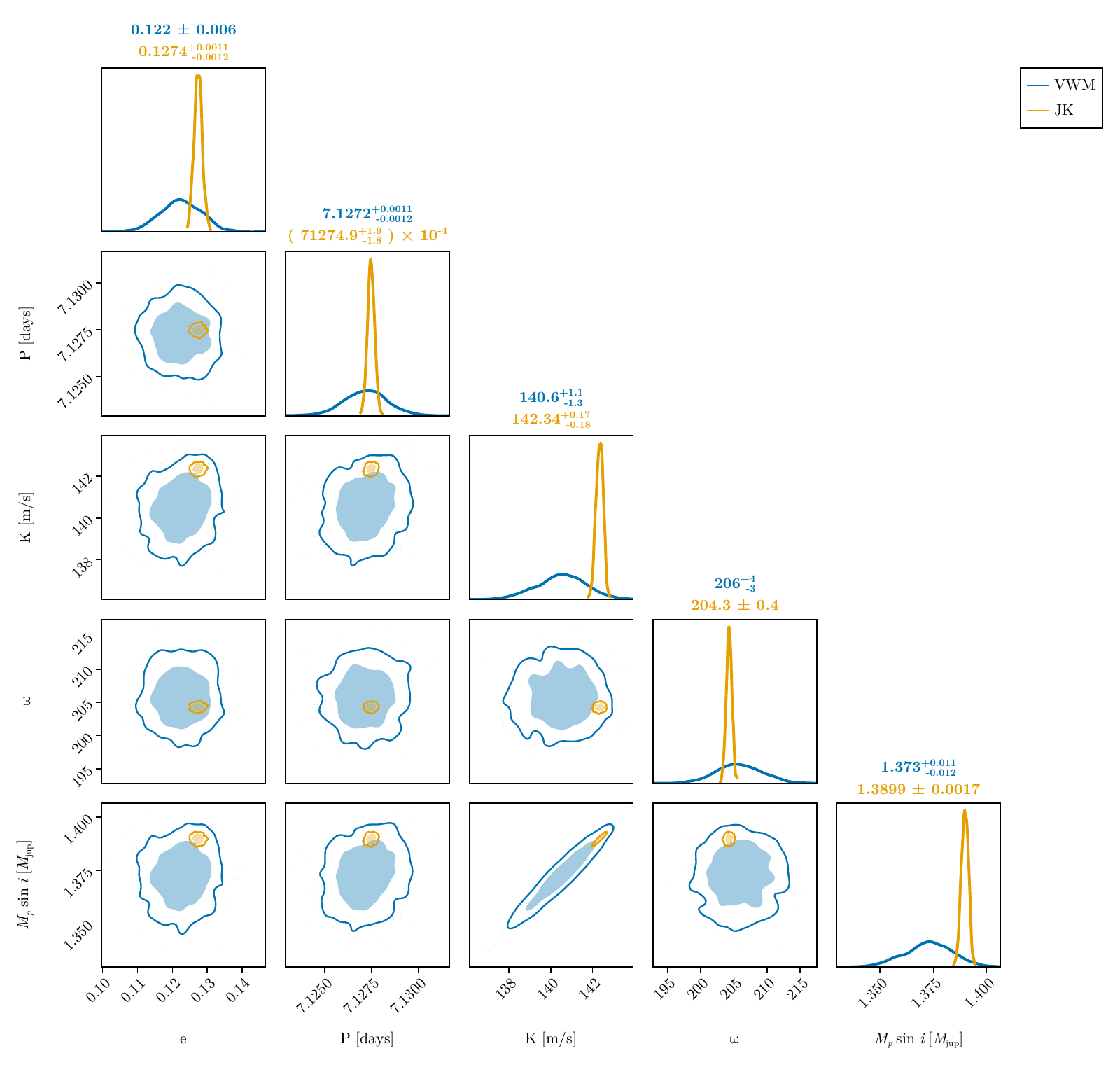}
    \caption{Joint posterior distributions for HD 217107.}
    \label{fig:pairplot_HD217107}
\end{figure*}

\begin{figure*}
    \centering
    \textbf{HD 3651}\par\medskip
    \includegraphics[width=\textwidth]{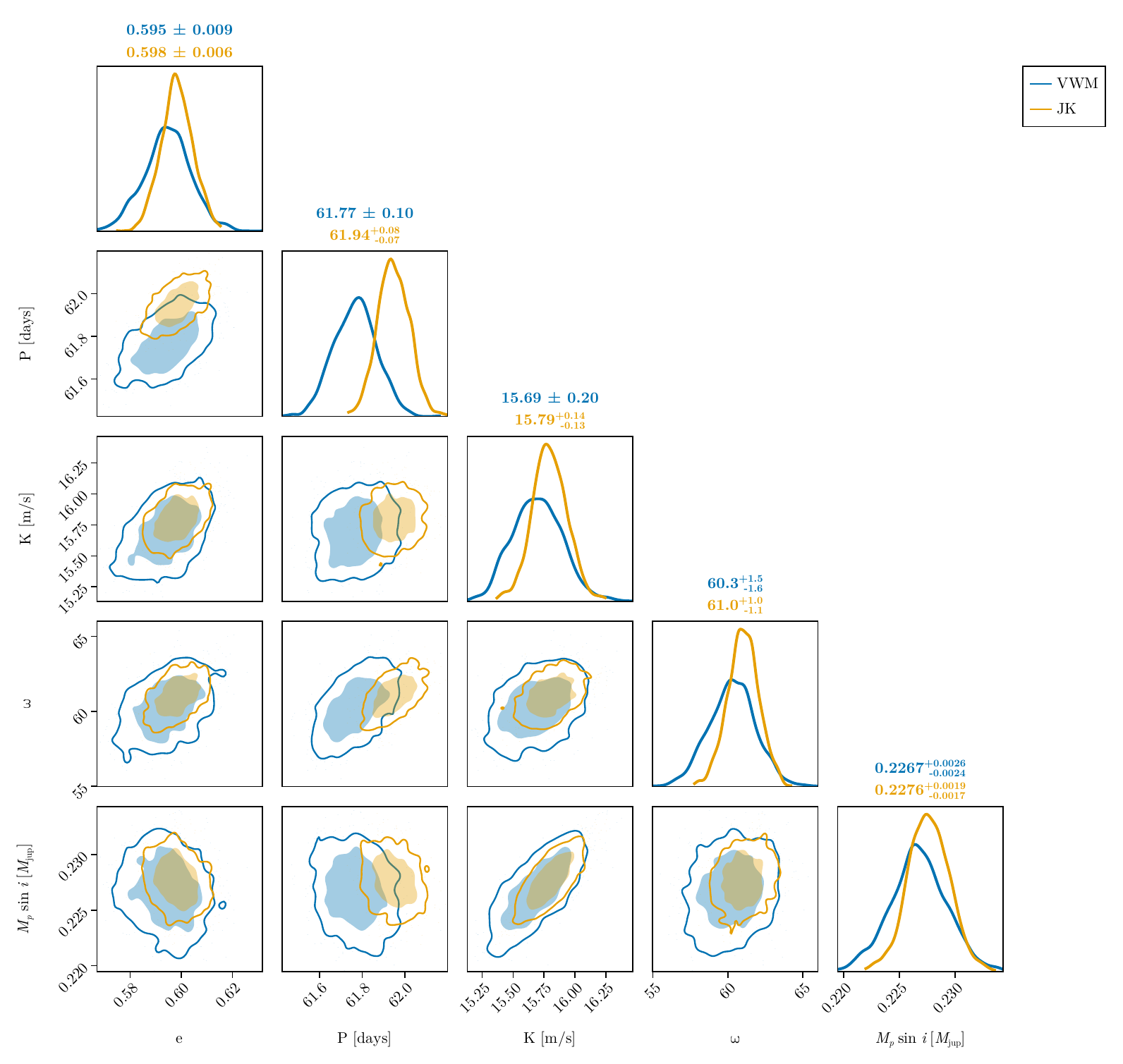}
    \caption{Joint posterior distributions for HD 3651.}
    \label{fig:pairplot_HD3651}
\end{figure*}

\begin{figure*}
    \centering
    \textbf{HD 86728}\par\medskip
    \includegraphics[width=\textwidth]{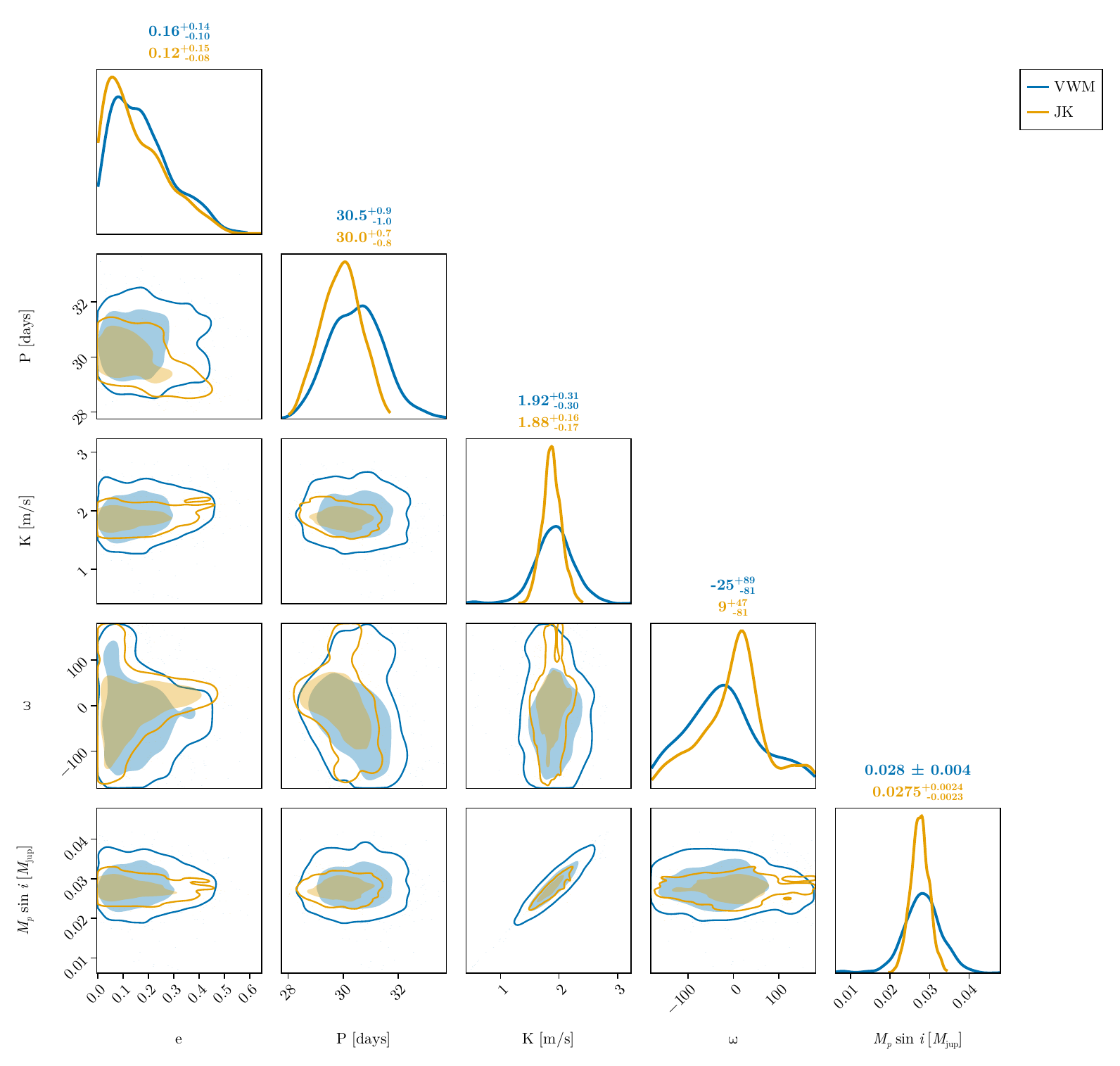}
    \caption{Joint posterior distributions for HD 86728.}
    \label{fig:pairplot_HD86728}
\end{figure*}

We show the VWM and JK posteriors for the model parameters ($K$, $P$, $e$, and $\omega$) and $M_p\sin{i}$ in Figures \ref{fig:pairplot_HD217107}, \ref{fig:pairplot_HD3651}, and \ref{fig:pairplot_HD86728}, and display the mean parameters and uncertainties from the JK analysis for each system in Table \ref{tab:data}. 

For all parameters, the posteriors for the JK case are more precise than the VWM. However, we note that we do not obtain better constraints with our methods than those in the literature for all parameters. For example, the period is much less constrained by our methods. We attribute this to the fact that our data contain fewer epochs over a shorter time-baseline than previous work. In the same way, we would expect the $M_p\sin{i}$ constraints to further improve when using more epochs.

\subsection{Results}\label{ssec:results}
Table \ref{tab:performance} shows the total compute time (upper) and the ratio of the $M_p\sin{i}$ posterior standard deviation of each method to the JK method (lower).

\begin{table}
\centering
\begin{tabular}{l|llll}
\tableline
System & VWM & OBO & JK &  \\
\tableline
HD 217107 & 35 sec & 13 min & 39 hr 13 min &  \\
 & 6.80 & 0.99 & -- &  \\
HD 3651 & 35 sec & 1 hr 34 min* & 7 hr 26 min &  \\
 & 1.5 & 0.97 & -- &  \\
HD 86728 & 30 sec & 18 min & 3 hr 20 min &  \\
 & 2.12 & 1.22 & -- & \\
 \tableline
\end{tabular}
\caption{The total compute time (upper) and the ratio of the $M_p\sin{i}$ posterior standard deviation of each method to the JK method (lower). E.g. For HD 217107, the OBO method takes 13 minutes to complete the analysis and the posterior is 0.97 times the width of the the JK method. *Note: Upon closer inspection, this system has a number of orders that produce chains with large correlation lengths. These few orders dominate the compute time, as implemented. Without the continued sampling (see $\S$ \ref{ssec:mcmc}), this method takes 16 minutes, which is more inline with the other systems.}
\label{tab:performance}
\end{table}

The VWM method is the fastest with a median time of 35 s. This is expected as the Keplerian orbit model is efficient to compute, and the data sets only contain tens of data points. Compared to JK, we see at least a 50\% increase in the posterior widths, up to a factor of approximately eight. 

The OBO method increases to tens of minutes, which is faster than simply N$\times$35 seconds, since the startup cost of running the code (e.g. pre-compiling, loading data, etc.) is the same as in the VWM case. We might also reduce the computation time by running individual orders in parallel. The improvement from the JK method is less stark here. However, we still see a 20\% larger posterior width for HD 86728. We also see a reduction of 1\% and 3\% for HD 217107 and HD 3651, respectively, compared to JK. 

Finally, the JK method introduces a larger computational cost -- anywhere from $\sim$3 hours to $\sim$2 days. This is likely due to the NUTS adaptation step taking a long time. If the estimate of the so-called 'mass-matrix' is poor, the numerical integration may take many steps, which drives up the computation time. 

As these are all distinct probabilistic models, the width of the posterior distribution is not the only consideration here, but also the bias. It is clear that we get a substantial precision gain going from VWM to OBO, with modest increases in computation time. In the HD 217107 and HD 3651 cases, OBO and JK are roughly within $2\sigma$ of the VWM measurements, and OBO is within $1\sigma$ of the JK measurement. For HD 86728, the discrepancy between OBO and JK is much larger -- needing to go out to $6\sigma$ in the JK distribution. Visually, all of the OBO distributions are biased to larger values than the JK (and VWM), with the discrepancy between OBO and JK seeming to increase as the semi-amplitude decreases. With this small of a sample size, it is unclear whether this is a real trend or an artifact of the particular systems we have analyzed.

We also note that these results are in agreement with the literature. For HD 217107 and HD 3651, the mean posteriors $M_p\sin{i}$ for all of our models are within $2\sigma$ of the reported measurements. For HD 86728, the VWM and JK methods are within $2\sigma$, while the OBO model is biased toward a larger minimum mass of about $5\sigma$. 

\subsection{Comparison to \cite{giovinazzi2025}}\label{ssec:mark}
A recent paper \citep{giovinazzi2025} analyzes the same set of NEID observations, together with additional RV and astrometry data, of HD 217107. In brief, they perform a joint fit to the RVs and astrometry, which allows them to constrain the mass and inclination of HD 217107 c, as well as refine the orbit for HD 217107 b \citep[see][$\S$4 for details]{giovinazzi2025}. The authors used 952 VWM analogous RV measurements spanning over 20 years. They find an $M_p\sin{i}$ of 1.371$^{+0.016}_{-0.020}$ $M_{\text{jup}}$, which is consistent with our JK measurement (1.389 $\pm$ 0.0017 $M_{\text{jup}}$) to just over $1\sigma$. 

\section{Conclusion} \label{sec:con}
Using a set of three systems, each with NEID multi-order radial velocity measurements, the Order-by-order (OBO; $\S$\ref{ssec:obo}) and Joint Keplerian (JK; $\S$\ref{ssec:jk}) methods seem promising for extracting more precise measurements of exoplanet orbital parameters. 

Analysis with the OBO method (Figure \ref{fig:obo_masses}) shows that there are common orders that exhibit correlated noise in wavelength (e.g. below $\sim$ 4000 Angstroms), which can bias the measurements of the planet's minimum mass. In addition, Figure \ref{fig:rms} shows that even orders with small RV uncertainties can lead to large posterior widths for $M_p\sin{i}$, which points to the fact that these errors are underestimated and could bias the overall VWM-derived RVs. This may be due to poorly fit CCFs or unaccounted for astrophysical noise. In future work, we plan to investigate the sources of this phenomenon.

The OBO method clearly performs well in measuring $M_p\sin{i}$ for the high SNR HD 217107 system -- producing nearly identical results to the more computationally expensive JK method. However, for the lower SNR cases, the OBO and JK results diverge. As noted in Section \ref{ssec:jk}, the JK method is the better motivated probabilistic model, as it takes into account all of the data at once and allows the achromatic signal present in all orders to be described by the same Keplerian orbit. In future work, we plan to examine in what SNR regimes the results of these two modeling techniques start to diverge.

From these three examples, we can see that the precision difference between VWM and JK seems to grow with $M_p\sin{i}$ (i.e., the amplitude of the barycentric RV signal). This points to the fact that VWM neglects even time-independent wavelength correlations in the data. In future work, we plan to explore more sophisticated noise models. For example, the modeling schemes presented here do not include Gaussian Processes, which are becoming commonly used in astronomy \citep{aigrain2023}. In certain conditions, it is possible to marginalize over the linear offset term $\gamma$, which would help reduce the size of the parameter space for the JK method, and perhaps speed up the computation time. We also plan to determine whether we can achieve similar precision increases for the other orbital parameters as with $M_p\sin{i}$ when more epochs are available.

Based on only the three systems analyzed here, we cannot definitively say which of these new methods are superior for any specific data set. This work focuses on a small subset of available EPRV data, with specific requirements such that we had as controlled an experiment as possible. However, even with limited examples, the JK method clearly improves upon VWM, for modest computational cost. In future work, we plan to apply these techniques to a larger array of data to include longer time-baselines, multi-instrument data, and multi-planet data sets -- aiming to provide a better determination of when our methods should be applied.

The modeling paradigms presented are straightforward to implement and can be applied generically to multi-wavelength data sets from instruments other than NEID. We choose to use the \texttt{Octofitter.jl} \texttt{Julia} package as they have already implemented a Keplerian orbit model, Gaussian likelihood functions, and MCMC sampling schemes. We provide the code used to perform the analyses in this paper as examples\footnote{\url{https://github.com/langfzac/obo-paper}}, and hope that this will enable the community to easily test out these methods on their own data sets.

\begin{acknowledgments}

We would like to thank the anonymous referee for thoughtful comments that have helped to improve this manuscript. 

NEID is funded by NASA through JPL contract 1547612 and the NEID Data Reduction Pipeline is funded through JPL contract 1644767. Data presented herein were obtained at the WIYN Observatory from telescope time allocated to NN-EXPLORE through the scientific partnership of the National Aeronautics and Space Administration, the National Science Foundation, and the National Optical Astronomy Observatory. Based on observations at Kitt Peak National Observatory, NSF’s NOIRLab, managed by the Association of Universities for Research in Astronomy (AURA) under a cooperative agreement with the National Science Foundation. The authors are honored to be permitted to conduct astronomical research on Iolkam Du\'ag (Kitt Peak), a mountain with particular significance to the Tohono O\'odham. We thank the NEID Queue Observers and WIYN Observing Associates for their skillful execution of our NEID observations.

Portions of this work were performed for the Jet Propulsion Laboratory, California Institute of Technology, sponsored by the United States Government under the Prime Contract 80NM0018D0004 between Caltech and NASA.

The Center for Exoplanets and Habitable Worlds is supported by Penn State and its Eberly College of Science.
The authors acknowledge the Penn State Institute for Computational and Data Science for providing computational resources and support that have contributed to the research results reported in this publication.

Work at the University of Pennsylvania is partially supported by NASA through JPL subcontract 1714025. ZL acknowledges the use of the University of Pennsylvania General Purpose Cluster, on which the computations in this paper were carried out. ZL is partially supported by the Kaufman Foundation through grant KA2023-136485.

\end{acknowledgments}

\software{
\texttt{Octofitter.jl} v6.0.1 \citep{thompson2023a},
\texttt{Makie.jl} \citep{Danisch2021},
\texttt{PairPlots.jl} \citep{thompson2023b},
\texttt{AdvancedHMC.jl} \citep{Xu2020},
\texttt{MCMCChains.jl} (\url{https://github.com/TuringLang/MCMCChains.jl})
}

\bibliography{main}{}
\bibliographystyle{aasjournal}


\end{document}